\documentclass[onecolumn]{elsart3}
\usepackage{graphicx}
\usepackage{natbib}
\journal{Neurocomputing}

\def\g#1{g_{\rm #1}}
\def\u#1{\,{\rm #1}}
\def\rem#1{}
\def\sbttl#1{: #1}
\def\NEURON{{\it NEURON}}
\def\mod#1{\;(\mathop{{\rm mod}}\nolimits\,{#1})}
\def\FLAG#1{ }
\begin{document}
\begin{frontmatter}
\title{Neurokinematic Modeling of Complex Swimming Patterns of the Larval Zebrafish }
\author[CIRCS,Physics]{Scott A. Hill},
\author[Biology]{Xiao-Ping Liu},
\author[Biology]{Melissa A. Borla},
\author[CIRCS,Physics]{Jorge V. Jos\'e\corauthref{cor}}, and
\author[CIRCS,Biology]{Donald M. O'Malley}
\address[CIRCS]{Center for Interdisciplinary Research in Complex Systems, Northeastern University, Boston MA USA}
\address[Physics]{Department of Physics, Northeastern University, Boston MA USA}
\address[Biology]{Department of Biology, Northeastern University, Boston MA USA}
\corauth[cor]{Corresponding author: jjv@neu.edu.}

\begin{abstract}
Larval zebrafish exhibit a variety of complex undulatory swimming
patterns.  This repertoire is controlled by the 300 neurons projecting
from brain into spinal cord. Understanding how descending control
signals shape the output of spinal circuits, however, is nontrivial.
We have therefore developed a segmental oscillator model (using
\NEURON) to investigate this system.  We found that adjusting the
strength of NMDA and glycinergic synapses enabled the generation of
oscillation (tail-beat) frequencies over the range exhibited in
different larval swim patterns. In addition, we developed a kinematic model to visualize the more complex axial bending patterns used
during prey capture.
\end{abstract}
\begin{keyword}zebrafish, swimming, locomotion, CPG, spinal cord
\end{keyword}
\end{frontmatter}
%==========================================================================
\section{Introduction}

The spinal cords of vertebrate animals contain segmental oscillators,
or central pattern generators (CPGs), that can produce rhythmic
movements.  The operations of these spinal CPGs are best understood in
lower vertebrates, such as lamprey and Xenopus, where they are used
for undulatory swimming \citep{Rob98,Buc99,Gri03}.  Control signals
descending from brainstem to spinal cord have also been studied
extensively in these and other lower vertebrates, such as goldfish and
zebrafish. Their lack of a corticospinal tract avoids a degree of
complexity that is present in mammals.  The functioning of descending
control systems in higher vertebrates has been difficult to
understand.  Studies of lower vertebrates should reveal conserved
principles by which these systems operate.  The larval zebrafish takes
vertebrate simplicity to an extreme: the decreased numbers of neurons
allows exact identification of many cell types in both brainstem and
spinal cord \citep{Kim82,Kim85,Liu88,Ber90}; this in turn provides
major experimental and modeling advantages.

One important aspect of lower vertebrate locomotion is the regulation
of swimming speed, which is often correlated with the frequency of
alternating left and right contractions of the axial muscles of the
trunk or tail, termed tail-beat frequency (TBF).  In a steady-swimming
fish TBF is generally equal to the oscillation frequency of the spinal
CPGs, so by understanding the modulation of CPG frequency in spinal
cord we can understand a major element of the control of swim speed.
In goldfish, lamprey, and other fishes, swim frequency can be
modulated by stimulation of the midbrain locomotor region or bath
application of NMDA; It is also known that serotonin, acetylcholine,
dopamine, and other neurotransmitters also influence the CPG's
oscillation frequency.  However, the identities and locations (in
brainstem or spinal cord) of the cells involved in swim frequency
control are unknown.  Larval zebrafish may provide insights into this
problem because they exhibit distinct swim patterns that span a broad
range of tail beat frequencies, ranging from 25 to 75 Hz
\citep{Bud00}.  A critical unknown is whether or not the distinct
larval swim patterns ({\it slow}, {\it burst}, and {\it capture}) are
controlled by distinct control systems and/or specific
neurotransmitters\citep{Bor02,OMa03}. Modeling the different types of
swimming behavior using a combined neural and kinematic model can shed
light onto the presence and relevance of different motor control
elements.

The functioning of spinal networks that underlie locomotion in fishes
and tadpoles has been extensively modeled \citep[see
e.g.][]{Rob90,Dal95,Dal03,Gri03}.  We created a zebrafish neural
model, based on previous Xenopus spinal network models, of the CPGs in
the larval zebrafish spinal cord, incorporating known properties of
the oscillators underlying swimming \citep{Tun02}.  The spinal
interneuron types in zebrafish \citep{Hal01} are likely homologous to
those in both Xenopus and lamprey \citep{Fet92}.  We explored the
control of tail-beat frequency, and found that by altering the
strengths of AMPA, NMDA and glycinergic-like synapses (all known to be
present in lower vertebrate spinal cords), we were able to generate
TBFs that spanned the range of speeds observed during burst and slow
swimming behaviors.  We also created a simple mechanical or
``kinematic'' model (that can be driven by the neural model) to
visualize how the spinal neural activity might be transformed into
larval behaviors.  Our ultimate goal in creating this neurokinematic
model is to introduce a tool for testing theories of descending motor
control in the larval zebrafish.

%==========================================================================
\section{Methods}
%\subsection{Neural Model}
%---------------------------------------------------------------------------------------------------------------------------------
%:figure: neural
\begin{figure}
\begin{center}\includegraphics[scale=0.6]{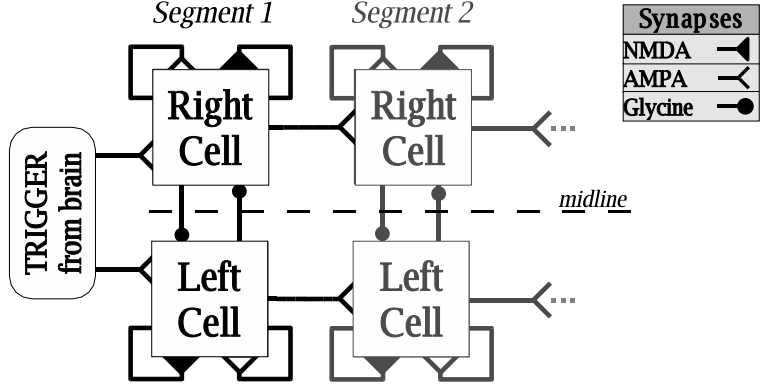}\end{center}
\caption{\label{fig-neural}Schematic diagram of the neural oscillator
  model.  The squares represent neurons, and each line indicates one
  of three types of synaptic connection: glycinergic (fast
  inhibitory), AMPA-like (fast excitatory), or NMDA-like (slow
  excitatory).  Only the first two segments of a 25-segment chain are
  shown.}
\end{figure}
%---------------------------------------------------------------------------------------------------------------------------------
Figure~\ref{fig-neural} shows the structure of our neural model, which is a simplification of the larval spinal circuitry \citep{Hal01} incorporating the minimal elements required to generate rhythmic, propagating alternating activity.  Each segmental oscillator consists of two neurons, with NMDA and AMPA-like autosynapses, which are connected by glycinergic synapses.  An
oscillator can be triggered with a single excitatory pulse to one
cell, followed by a single pulse to the other cell several
milliseconds later (7~ms in our simulations).  Individual oscillators
are connected into a 25-segment chain (corresponding to the
approximately 25 segments in the zebrafish spinal cord), with
nearest-neighbor descending excitatory synaptic connections; the
entire chain can be started merely by triggering the head segment.

We use cells with the standard Hodgkin-Huxley channels provided by the
\NEURON\ software package: persistent potassium ($\bar
\g{K}=0.036\u{S/cm^2}$, $E_{\rm K}=-77\u{mV}$), transient sodium
($\bar \g{Na}=0.12\u{S/cm^2}$, $E_{\rm Na}=50\u{mV}$), and leak
($\g{L}=0.0003\u{S/cm^2}$, $E_{\rm L}=-54.3\u{mV}$) channels.  Our
synapses are modeled using the difference of two exponentials, with
rise time $\tau_1(=1\u{ms})$ and fall time $\tau_2$.  See
\citet[p. 182]{DA01} for details.  The three types of synapses we used
were AMPA ($\tau_2=6\u{ms}$, $E=0\u{mV}$), NMDA ($\tau_2=80\u{ms}$,
$E=60\u{mV}$), and glycine ($\tau_2=2\u{ms}$, $E=-80\u{mV}$).

%\subsection{Kinematic Model}
Our kinematic model is meant to translate the neural signals from the
spinal circuitry into observable kinematic behaviors.  To do so, we use
a relatively simple transformation of putative neural output into
effects on the radius of curvature of a line segment representing the
trunk of the larva.  This model does not take into account the full
physics (e.g., elasticity and hydrodynamics) of the situation;
nevertheless, it does approximate the observed axial kinematics
recorded experimentally with a high-speed camera.

Suppose that each segment $x$ is receiving a neural signal $F_s(x,t)$
at time $t$ from the spinal circuitry: $F_s>0$ for a signal to the
right side of the segment, $F_s<0$ for a signal to the left.  We
suppose that this signal is integrated through exponential synapses,
so that the signal passing to the muscles is
\begin{equation}
F_m(x,t)=\int^t_{-\infty} F_s(x,t')[e^{(t'-t)/\tau_2}-e^{(t'-t)/\tau_1}] dt',
\end{equation}
where $\tau_1$ and $\tau_2$ are the growth and decay time constants of
the synapse (we use $\tau_1=6\u{ms}$ and $\tau_2=8\u{ms}$ in our
calculations.)  Assuming that the muscle contracts linearly as a
function of $F_m$, it follows that the radius of curvature of segment
$x$ is
\begin{equation}
\label{eq-radius}
R(x,t)=W(x)/F_m(x,t)
\end{equation}
$W(x)$ is a function describing the stiffness, or resistance to
bending, of segment $x$.  It takes into account the width of the
body (the tail is more flexible, and bends more than the rostral trunk).

We can feed the output of our neural model directly into this system,
but it is also useful to introduce a (more) artificial signal.  We
build this signal out of three components: an oscillatory signal
$F_{osc}(x,t)$, which is a series of delta functions propagating
caudally; a bending signal $F_{bend}$, which is a tonic signal applied
to one side of all segments at once; and a rostral stiffening signal,
which reduces the signal to the first $x_{inh}$ rostral segments by a
factor $f_{inh}$.  The first two pieces are responsible for swimming
and turning, respectively.  The stiffening signal is seen in larvae
during prey capture, where the fish keeps its head relatively still
while adjusting its orientation with its tail.  We can write our
artificial neural signal as
\begin{equation}
\label{Fs}
F_s(x,t)=(F_{osc}(x,t)+f_{bend})\times\cases{f_{inh},&$x\le x_{inh}$\cr 1,&$x > x_{inh}$},
\end{equation}
where $f_{bend}$ is a constant and
\begin{equation}
\label{Fosc}
F_{osc}(x,t)=f_{osc}\cases{1,&$x\equiv 2\pi\nu t \mod{\lambda}$\cr
                    -1,&$x\equiv2\pi\nu t+\lambda/2 \mod{\lambda}$\cr
                    0, & otherwise}
\end{equation}
The parameter $\nu$ is the tail-beat frequency, and $\lambda$ is the
length (in number of segments) of the resulting wave which propagates
down the fish.
%==========================================================================
\section{Results and Discussion}
%---------------------------------------------------------------------------------------------------------------------------------
%:figure: windup
\begin{figure}
\begin{center}\includegraphics[width=3.5in]{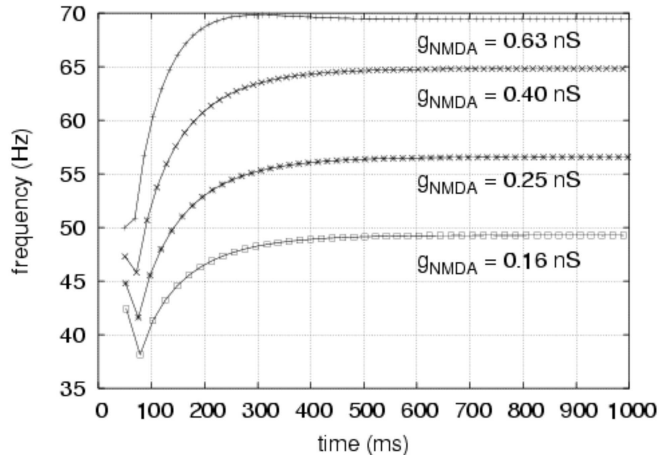}
\end{center}
\caption{\label{fig-windup}Instantaneous TBF in a single segmental
  oscillator, for several values of $\g{NMDA}$.  Each point represents
  the iTBF for the left neuron; the corresponding points from the
  right side gives near-identical results.  Oscillators approach a
  steady-state frequency only after an initial ``wind-up'' period.}
\end{figure}
%---------------------------------------------------------------------------------------------------------------------------------
We began by studying a single segment of the neural model shown in
Fig.~\ref{fig-neural}.  A transient trigger pulse to this model
initiates sustained alternating activity, mimicking Xenopus spinal
cord, where a transient stimulus produces a sustained bout of swimming
\citep{Rob90}.  The first larval-zebrafish specific task was to
generate the range of tail-beat frequencies (TBFs) used in different
swimming behaviors.  The frequency of oscillation was calculated on a
half-cycle basis \citep[termed ``instantaneous'' TBF in ][]{Bor02}.
Figure~\ref{fig-windup} shows how instantaneous TBF varies with time,
with each trace representing the frequency profile for a specific
value of the NMDA synaptic conductance.  In all cases, there is an
initial transient period, 15 to 20 cycles long, where the TBF
increases by about a third before settling into an indefinite steady
state with constant frequency.  We subsequently re-inspected
behavioral sequences, and found varying degrees of wind-up in some,
but not all, experimentally recorded swim bouts \citep{Bor02}.  It is
unknown whether the biological and simulational wind-up behaviors are
related, but the phenomenon lasts much longer in the model than in the
lab.

%---------------------------------------------------------------------------------------------------------------------------------
%:figure: range
\begin{figure}
\begin{center}
\includegraphics[width=3in]{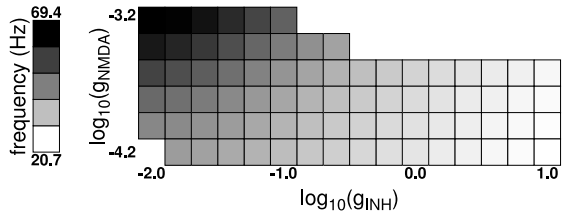}
\caption{\label{fig-range}A density plot showing the steady-state
  frequency of a single oscillator as we modify the synaptic
  conductances $\g{NMDA}$ and $\g{INH}$, fixing $\g{AMPA}=10^{-4}$.
  All conductances are in $\u{\mu S}$, and all frequencies are in Hz.
  The missing regions in the upper-right and lower-left corners
  represent non-oscillatory states.}
\end{center}
\end{figure}
%---------------------------------------------------------------------------------------------------------------------------------
To evaluate the range of oscillator or tail-beat frequencies that
might be generated with this simple model, we tested different
combinations of synaptic strengths (conductances).  Varying the
strengths of the NMDA and glycinergic synapses was found to alter the
steady-state oscillator frequency (Fig.~\ref{fig-range}): increasing
the NMDA conductance increased TBF, whereas increasing the glycinergic
conductance decreased it.  The AMPA conductance can also affect TBF,
but to a lesser extent.  Different combinations of conductances could
give rise to the entire range of TBFs observed in different larval
swim patterns, from the slow swim (25 to 40 Hz) to burst swim
\citep[45 to 75 Hz;][]{Bud00}.  This is just one potential means of
varying oscillator frequency (or TBF), implemented in a reduced system
(i.e. a two-cell model), but it illustrates how a minimal model of
larval zebrafish spinal cord, with just a few essential conductances,
can give rise to a range of outputs relevant to the larval behaviors.
In this model, each crossed inhibitory signal results in one
post-inhibitory rebound firing of a single action potential, but to
more completely capture slow and burst swims we will need to
incorporate mechanisms that regulate the strength of output of the
motoneuron pools.  \FLAG{Also, larval zebrafish swim bouts last for
only a few cycles, so a more complete model would need to incorporate
network elements or channel properties that result in fairly rapid
cessation of swimming.}

%---------------------------------------------------------------------------------------------------------------------------------
%:figure: raster
\begin{figure}
\begin{center}\includegraphics[width=3in]{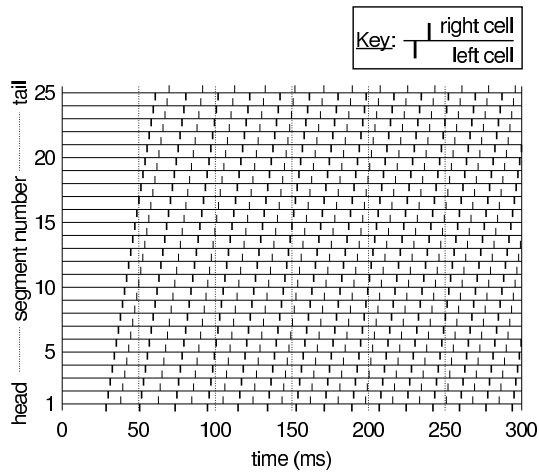}
\end{center}
\caption{\label{fig-raster}Rastergram for a chain of 25 segmental
  oscillators.  Each horizontal line corresponds to a single
  oscillator; ticks drawn below each line are for action potentials on
  the left, and above for those on the right.  Parameters:
  $\g{AMPA}=10^{-4}\u{\mu S}$, $\g{NMDA}=6\times 10^{-4}\u{\mu S}$,
  $\g{INH}=10^{-2}\u{\mu S}$, $\g{descending}=10^{-2}\u{\mu S}$.  }
\end{figure}
%---------------------------------------------------------------------------------------------------------------------------------
We next extended the model to a 25-segment chain of oscillators,
creating a kind of artificial spinal cord that should facilitate
quantitative analyses of the influences of descending signals on
spinal network activity.  The firing pattern of the model spinal cord
is illustrated in Fig.~\ref{fig-raster}.  The firing of segment number
1 shows an alternating left-right pattern over the duration of the
simulation.  In successively more caudal segments, the firing is
delayed, matching the general pattern of undulatory swimming.  As in
the single segment model, each segment along the chain showed a
frequency wind-up period.  In this model a fixed intersegmental
time-delay (nominally a synaptic delay) was used to establish the
phase relationship between segments, but a more realistic model might
include ascending and descending connections of varying strengths and
lengths, as in lamprey (\citealt{Kot99}, and see \citealt{Dal03}).
\FLAG{While lampreys tend to have a constant inter-segmental phase
relationship at different swimming speeds, our observations suggest
that larval zebrafish swim patterns show different wavelengths of
axial bending, and therefore a varying phase relationship.  Our model
will therefore need descending controls that are able to adjust this
phase relationship appropriately for different swim patterns.}

%---------------------------------------------------------------------------------------------------------------------------------
%:figure: real
\begin{figure}
\includegraphics[height=1.05in]{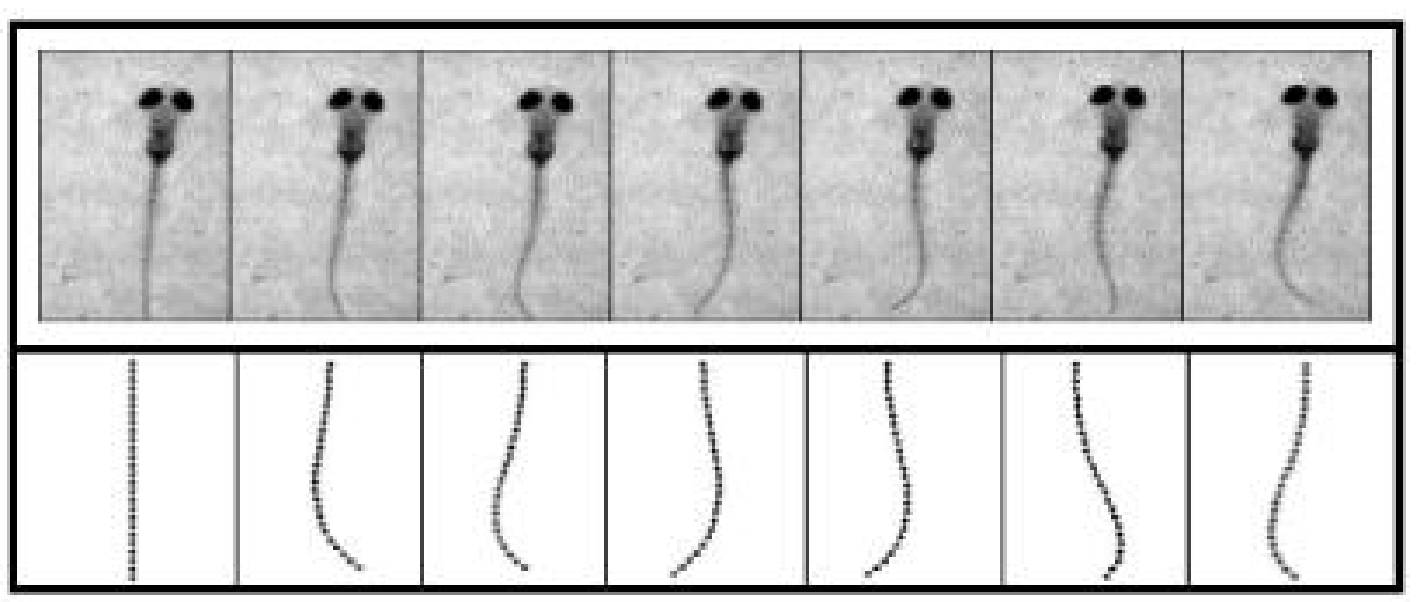}
\quad
\vbox{
\includegraphics[height=0.5in]{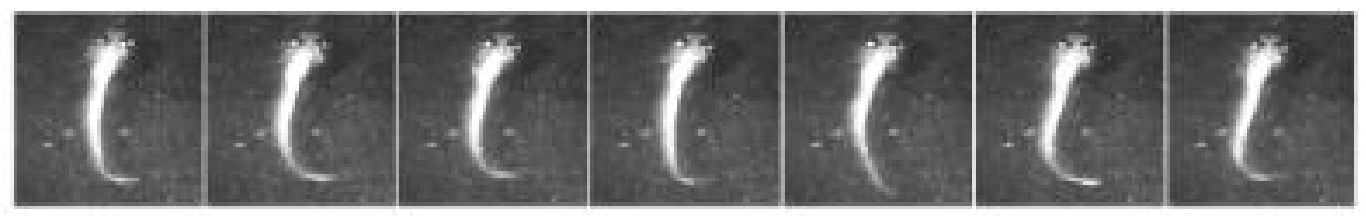}\par
\includegraphics[height=0.5in]{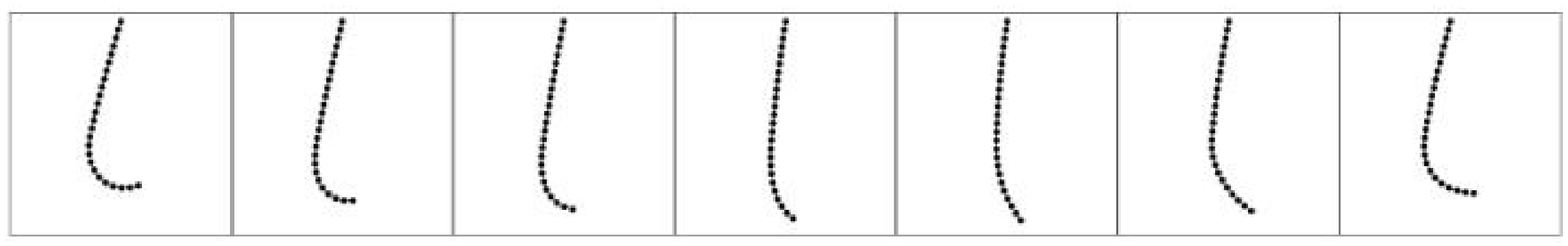}
}
\caption{\label{fig-real}Comparison of a 7-day old larval zebrafish
  with our kinematic model during a slow swim (left) and a J-turn (right).}
\end{figure}
%---------------------------------------------------------------------------------------------------------------------------------
The kinematic model of the larval trunk is a complementary tool for
exploring theories of descending motor control.  Preliminary work
\citep{Hil03} assumed that the fish was equally flexible from trunk to
tail ($W(x)=1$ in Eq.~\ref{eq-radius}), with partial success.  By
accounting for the increased flexibility of the tail (e.g. by making
$W(x)$ linear), however, we can better simulate the trunk kinematics
observed during slow swims and J-turns
(Fig.~\ref{fig-real}).  J-turns are unique locomotive maneuvers that
contribute to the larval prey-capture behavior \citep{BoO02}.  They
require repetitive, asymmetric and far-caudal contractions of axial
musculature.  This might in principle be achieved by sending an
asymmetric excitatory brainstem signal exclusively to far-caudal
spinal cord; however, a survey of the spinal outputs of zebrafish
reticulospinal neurons revealed no neurons with the requisite
arborization pattern \citep{Gah03}.  Alternatively, neurons that
selectively arborize in rostral spinal cord (which were observed)
might be activated bilaterally to stiffen the rostral musculature. In
conjunction with such signals, other neurons that arborize along the
entire length of spinal cord could then generate far-caudal
contraction.  In our model, this is implemented as a bilateral
``inhibitory'' rostral signal, but this is kinematically equivalent to
stiffening the rostral end of the larva by bilateral excitation of
rostral motoneurons.

\medskip
\noindent{\it Future Directions}
\medskip

As we extend the neural model's capabilities to produce dynamically
varying bend amplitudes, we hope to produce an increasingly realistic
``artificial spinal cord'' that can be used to test ideas of how the
larval locomotive repertoire is generated. The kinematic model is
useful here because it provides a first approximation of how complex
neural model outputs might affect axial kinematics.  Further
extensions of this model should incorporate known constraints of the
larval CNS, such as the diversity of brainstem and spinal systems
involved in swimming and turning behaviors \citep{OMa96,Hal01,Rit01}, and
the wide distribution of activity during escape behaviors
\citep{Bos01,Gah02}.\FLAG{and the resistance of this activity to the
disruptive effects of laser-ablations \citep{Gah01}.}  By incorporating
such constraints, the combined neurokinematic model should become
increasingly useful in generating experimentally-testable hypotheses
of descending and spinal control of vertebrate locomotion.

\section{Acknowledgements}
Support for this work was provided by NIH-NS37789 (DMO) and by the
Center for Interdisciplinary Research in Complex Systems (CIRCS) at
Northeastern University (SAH, JVJ).
%==========================================================================
%:Bibliography

\end{document}